# ESTIMATE OF THE TRUNCATION ERROR OF FINITE VOLUME DISCRETISATION OF THE NAVIER-STOKES EQUATIONS ON COLOCATED GRIDS.

**Alexandros Syrakos, Apostolos Goulas**
Department of Mechanical Engineering, Aristotle University of Thessaloniki, Greece.

**Abstract**

A methodology is proposed for the calculation of the truncation error of finite volume discretisations of the incompressible Navier – Stokes equations on colocated grids. The truncation error is estimated by restricting the solution obtained on a given grid to a coarser grid and calculating the image of the discrete Navier – Stokes operator of the coarse grid on the restricted velocity and pressure field. The proposed methodology is not a new concept but its application to colocated finite volume discretisations of the incompressible Navier – Stokes equations is made possible by the introduction of a variant of the momentum interpolation technique for mass fluxes where the pressure-part of the mass fluxes is not dependent on the coefficients of the linearised momentum equations. The theory presented is supported by a number of numerical experiments. The methodology is developed for two-dimensional flows, but extension to three-dimensional cases should not pose problems.

## 1. Introduction

Finite volume methods, and especially those of $2^{nd}$-order accuracy, are very popular for the solution of the Navier-Stokes equations because, by today's standards, they offer acceptable accuracy on reasonably dense grids while being easy to implement. The truncation error is the measure of the discrepancy between the discrete system which arises from application of the finite volume methodology and the original integral-differential equations.

For structured grids several truncation error estimators have been proposed for particular discretisation schemes, for example in [1], [2] and [3]. They express the leading term of the truncation error in terms of the derivatives of the flow variables and the geometry of the grid. These estimators are useful but they have the disadvantage that they are different for each discretisation scheme, and that they apply in the case of structured grids which have been constructed from distributions of dimensionless variables with continuous derivatives. For more general cases, a number of truncation error indicators have been proposed, as in [4], [5] or [6], that is, quantities which resemble the truncation error and sometimes have the same units, and are likely to be large in regions where the truncation error is large. However, these quantities may fail to capture certain parts of the truncation error, for example the skewness-induced part if the indicator is constructed from a one-dimensional analysis as in [4]. Besides, an estimate would be more useful rather than an indication.

Multigrid solution methods which use the full approximation scheme (FAS) automatically provide a quantity, the relative truncation error between the finest and the immediately coarser grid, which can easily be converted into a truncation error estimate if the order of the discretisation is known – see [7], [8], [9]. In fact, the truncation error estimator can be used independently of the multigrid procedure. All that is required is a solution on a given grid, and a coarser grid which is similar to the fine grid. The equations need not be solved on the coarse grid. This inspired the present authors to implement this estimator in the case of the finite volume discretisation of the incompressible Navier-Stokes equations on colocated grids.

## 2. Finite volume discretisation: Basic principles and notation

Here it will be useful to introduce some notation, which is similar to that used in [9]. The domain is decomposed into a finite number of control volumes (CVs) using a grid. A grid is denoted by a letter such as $h$, which will also be interpreted as the distribution of the grid spacing in the domain[†]. Therefore grid $a \cdot h$ is such that it's spacing at each location equals $a$ times the spacing of grid $h$ at the same location. Given a grid $h$, such a process of obtaining grid $a \cdot h$ will be referred to as *systematic refinement* if $a < 1$, or *systematic unrefinement* if $a > 1$. The set of all CVs of a grid $h$ is denoted as $V_h$.

The set of all points in the region where the partial differential equation is defined will be denoted by $\Omega$, and $G(\Omega)$ will denote the set of functions which are defined on $\Omega$. Analogously $\Omega_h \subset \Omega$ will denote the set of centroids of the CVs of grid $h$ and $G(\Omega_h)$ will denote the set of all functions defined on $\Omega_h$ (grid functions). Also if $\phi \in G(\Omega)$ then $\phi_h \in G(\Omega_h)$ will denote the grid function such that $\phi_h(x) = \phi(x)$ for all $x \in \Omega_h$. Letters in bold italic such as $x$ refer to position vectors in space. Also $\phi_{h,P}$ or $(\phi_h)_P$ is the $P$-th component of the grid function $\phi_h$, that is the value of $\phi_h$ at the centroid of CV $P$. The operator which samples the function $\phi$ at the CV centroids to return the grid function $\phi_h$ is given as $I_0^h : G(\Omega) \to G(\Omega_h)$, i.e. $I_0^h \phi = \phi_h \Leftrightarrow \phi_h(x) = \phi(x)$ for all $x \in \Omega_h$. Since grid functions are usually used to represent functions of continuous space one can define the inverse operator $I_h^0 : G(\Omega_h) \to G(\Omega)$ such that $I_h^0 \phi_h = \phi \Rightarrow \phi(x) = \phi_h(x)$ for all $x \in \Omega_h$. For the rest of the points $x \notin \Omega_h$, $\phi(x)$ will assume a value determined by a suitable interpolation, so $I_h^0$ is not unique. By similar reasoning a grid function may be transferred from a grid $h$, say, to a grid $k$ by the operator $I_h^k : G(\Omega_h) \to G(\Omega_k)$, $I_h^k = I_0^k \cdot I_h^0$. Again a suitable interpolation must be chosen.

Suppose a differential operator $N$ which acts on a function $\phi$ and returns a function $N\phi$. The finite volume method approximates the integral of the operator image over each CV by an algebraic expression. If $P$ is a CV covering a volume $\Delta\Omega_P$ then the finite volume method starts by deriving a relation of the form:

$$\frac{1}{\Delta\Omega_P} \int_{\Delta\Omega_P} N\phi \, d\Omega = \left(N_h \phi_h\right)_P + \tau_{h,P} \qquad (2.1)$$

where $N_h$ is an algebraic operator which approximates the average of $N$ over each CV and $\tau_h$ is the *truncation error* associated with $N_h$. The discretisation should be such that the truncation error tends to zero as $h \to 0$. In this case the left hand side of (2.1) tends to the value of $N\phi$ at the centroid of $P$, and since $\tau_h \to 0$ so does $(N_h \phi_h)_P$ due to (2.1). The smaller $\tau_{h,P}$ the better $N_h$ approximates the average of $N$ at CV $P$. Equation (2.1) is written in a form which aids theoretical understanding but in practice finite volume methods usually construct an algebraic expression which is equivalent to $\Delta\Omega \cdot N_h$ in an effort to approximate $\int_{\Delta\Omega_P} N\phi \, d\Omega$.

To solve the differential equation $N\phi = b$, where $b$ is a known function, by the finite volume method, the equation is first integrated over each CV giving $\int_{\Delta\Omega_P} N\phi \, d\Omega = \int_{\Delta\Omega_P} b \, d\Omega$ for each $P \in V_h$, and then substituting (2.1) for the left hand side one obtains:

$$\left(N_h \phi_h\right)_P + \tau_{h,P} = \frac{1}{\Delta\Omega_P} \int_{\Delta\Omega_P} b \, d\Omega \qquad (2.2)$$

---

[†] The grid spacing need not be a physical spacing. For example, for structured grids it may be defined as the spacing in the computational domain (as opposed to the physical domain). What is important is that it be defined so that relations of the form (2.5) - (2.7) may be derived.

Equation (2.2) is exact and so if one was able to solve it one would obtain $\phi_h$, the exact values of $\phi$ at the centroids of the CVs. Unfortunately, this is not possible since $\tau_h$ is not known. Instead one makes the assumption that the truncation error is small enough such that dropping it would not change the solution of the system significantly. Thus instead of (2.2), the following system is solved over each CV *P*:

$$\left(N_h \phi_h^*\right)_P = \frac{1}{\Delta\Omega_P} \int_{\Delta\Omega_P} b \, d\Omega \qquad (2.3)$$

The solution $\phi_h^*$ of the system (2.3) is not the same as the exact solution $\phi_h$. It differs by the *discretisation error* $\varepsilon_h = \phi_h - \phi_h^*$. As the grid is systematically refined and $h \to 0$ the truncation error will tend to zero and the systems (2.2) and (2.3) will tend to become equivalent. Therefore as $h \to 0 \Rightarrow \phi_h^* \to \phi_h$ and $\varepsilon_h \to 0$.

An analytic expression for $\tau_h$ can be derived as $N_h$ is constructed from $N$ using Taylor series. The truncation error for CV *P* will be of the form:

$$\tau_{h,P} = \sum_{k=p}^{\infty}\left(\sum_{n=1}^{1+nb} c_{k,n} \cdot h_n^k\right) \qquad (2.4)$$

for some $p \geq 1$, where *nb* is the number of neighbours of CV *P* which participate in its finite difference stencil and $h_n$ is the characteristic size of each of these neighbours, including *P* itself. The coefficients $c_{k,n}$ will be functions of the derivatives of $\phi$ in the vicinity of *P*. If the grid is refined systematically then the characteristic sizes of the neighbours of *P* will be proportional to the characteristic size *h* of *P* itself. In this case (2.4) may be written as:

$$\tau_{h,P} = \sum_{k=p}^{\infty} c_k \cdot h^k \qquad (2.5)$$

Through systematic refinement the space originally occupied by CV *P* will become occupied by more CVs. However, as the refinement is systematic the truncation error for these new CVs will be given by the same formula as for *P*. In addition if the derivatives of $\phi$ vary continuously and the grid spacing *h* is small enough then the coefficients $c_k$ will not be very different for the new CVs than for the original CV *P*. The change in the magnitude of the truncation error will therefore be mostly due to the reduction in grid spacing *h*. As systematic refinement progresses the terms $c_k \cdot h^k$ with $k > p$ of (2.5) will eventually become negligible compared to the term $c_p \cdot h^p$ and the truncation error will be reduced almost proportionally to $h^p$. The discretisation scheme $N_h$ is characterised as *p*-th order accurate.

For linear operators it can be shown that systematic refinement causes the discretisation error to reduce at the same rate as the truncation error. The same has been demonstrated experimentally for the Navier-Stokes operator by many researchers – see e.g. [10], [11].

The above discussion is summarised by the following relations, which hold for a *p*-th order accurate discretisation scheme, and which state that through systematic refinement the truncation and discretisation errors, treated as functions of the continuous space $\Omega$, retain their shape but their magnitude tends to become proportional to $h^p$:

$$I_h^0 \tau_h \in O(h^p) \qquad (2.6)$$

$$I_h^0 \varepsilon_h \in O(h^p) \qquad (2.7)$$

## 3. Truncation error estimate

The truncation error estimator, which is the one used in [8] and [9], will now be briefly described. To estimate $\tau_h$ on grid $h$, this estimator considers the same discretisation scheme on a systematically coarser grid, say $2h$. If, for brevity, one defines the grid function $b_h$ whose $P$-th component equals $(\int_{\Delta\Omega_P} b\, d\Omega) / \Delta\Omega_P$ (i.e. the right-hand side of (2.2)), then on grid $2h$ the relations which correspond to (2.2) and (2.3) are:

$$N_{2h}\phi_{2h} + \tau_{2h} = b_{2h} \tag{3.1}$$

$$N_{2h}\phi_{2h}^* = b_{2h} \tag{3.2}$$

From (2.6) and (2.7) it is deduced that:

$$I_{2h}^0 \tau_{2h} \approx 2^p \cdot I_h^0 \tau_h \tag{3.3}$$

$$I_{2h}^0 \varepsilon_{2h} \approx 2^p \cdot I_h^0 \varepsilon_h \tag{3.4}$$

A simple estimate of the truncation error begins by trying to estimate $\tau_{2h}$. Equation (3.1) cannot be used to calculate $\tau_{2h}$ because the exact solution $\phi_{2h}$ is not known. However, since the solution $\phi_h^*$ on grid $h$ is more accurate than $\phi_{2h}^*$, one may use it to approximate the exact solution. Therefore an estimate for $\tau_{2h}$ comes by substituting $I_h^{2h}\phi_h^*$ instead of $\phi_{2h}$ in (3.1):

$$\tau_{2h} \approx b_{2h} - N_{2h}\left(I_h^{2h}\phi_h^*\right) \equiv \tau_{2h}^h \tag{3.5}$$

The quantity $\tau_{2h}^h = b_{2h} - N_{2h}(I_h^{2h}\phi_h^*)$ is called the relative truncation error of grid $2h$ with respect to grid $h$ (it is used in the context of FAS multigrid methods). It can be readily calculated given the two grids $h$ and $2h$, and the solution $\phi_h^*$ on the fine grid. It is also dependent on the restriction operator $I_h^{2h}$. Adding $\tau_{2h}^h$ to the left hand side of (3.2) makes the solution of this system equal to the fine grid solution $I_h^{2h}\phi_h^*$, just like adding $\tau_{2h}$ to the left hand side of (3.2) makes the solution of this system equal to the exact solution $\phi_{2h}$.

Since $\tau_{2h}^h$ is an approximation to $\tau_{2h}$, one can use (3.3) to obtain an approximation for $\tau_h$. However, a more accurate estimate is possible. If $N_{2h}'(\phi_{2h}^*)$ is the jacobian matrix of the discrete operator $N_{2h}$ at $\phi_{2h}^*$ and the grid $2h$ is fine enough such that the differences between the functions $\phi_{2h}$, $I_h^{2h}\phi_h^*$ and $\phi_{2h}^*$ are small enough then the following hold:

$$N_{2h}\phi_{2h} \approx N_{2h}\phi_{2h}^* + N_{2h}'\left(\phi_{2h}^*\right)\cdot\left(\phi_{2h}-\phi_{2h}^*\right) \Rightarrow N_{2h}'\left(\phi_{2h}^*\right)\cdot\left(\phi_{2h}-\phi_{2h}^*\right) \approx -\tau_{2h} \tag{3.6}$$

$$N_{2h}I_h^{2h}\phi_h^* \approx N_{2h}\phi_{2h}^* + N_{2h}'\left(\phi_{2h}^*\right)\cdot\left(I_h^{2h}\phi_h^*-\phi_{2h}^*\right) \Rightarrow N_{2h}'\left(\phi_{2h}^*\right)\cdot\left(I_h^{2h}\phi_h^*-\phi_{2h}^*\right) \approx -\tau_{2h}^h \tag{3.7}$$

The second approximate relation of (3.6) derives from the first one due to (3.1) and (3.2), and similarly the second approximate relation of (3.7) derives from the first one due to the definition of $\tau_{2h}^h$ and (3.2). Using the fact that $\varepsilon_{2h} = \phi_{2h} - \phi_{2h}^*$ in (3.6), and the fact that $I_h^{2h}\phi_h^* - \phi_{2h}^* = (\phi_{2h} - \phi_{2h}^*) - (\phi_{2h} - I_h^{2h}\phi_h^*) \approx \varepsilon_{2h} - I_h^{2h}\varepsilon_h$ in (3.7), there result respectively:

$$N_{2h}'\left(\phi_{2h}^*\right)\cdot\varepsilon_{2h} \approx -\tau_{2h} \tag{3.8}$$

$$N_{2h}'\left(\phi_{2h}^*\right)\cdot\left(\varepsilon_{2h}-I_h^{2h}\varepsilon_h\right) \approx -\tau_{2h}^h \tag{3.9}$$

But (3.9) can change further by deducing from (3.4) that $I_h^{2h}\varepsilon_h \approx \varepsilon_{2h} / 2^p$ so that $\varepsilon_{2h} - I_h^{2h}\varepsilon_h \approx [(2^p -1) / 2^p]\cdot\varepsilon_{2h}$. Therefore (3.9) gives:

$$\frac{2^p - 1}{2^p} N'_{2h}\left(\phi^*_{2h}\right) \cdot \varepsilon_{2h} \approx -\tau^h_{2h} \tag{3.10}$$

Comparing (3.8) and (3.10) one gets:

$$\tau_{2h} \approx \frac{2^p}{2^p - 1} \tau^h_{2h} \tag{3.11}$$

and using (3.3) one arrives at the final truncation error estimator:

$$\tau_h \approx \frac{1}{2^p - 1} I^h_{2h} \tau^h_{2h} \tag{3.12}$$

Summarising, to estimate the truncation error: first solve the system on grid $h$, second restrict the solution to grid $2h$, third calculate the relative truncation error by (3.5), and finally apply (3.12). For this estimate to work it is crucial that the operator $N_{2h}$ is constructed using the same discretisation schemes as $N_h$. Actually it is not necessary to use grid $2h$, any multiple $r \cdot h$ will do and (3.12) holds with $r$ in place of 2. Also to ensure that the errors introduced in the restriction of the fine grid solution do not spoil the truncation error estimate it would be a good idea to use in (3.5) a restriction operator $I_h^{2h}$ of order higher than the order $p$ of the discretisation. This will ensure that as the grid is refined the error introduced by the restriction operator will eventually become negligible compared to the truncation error. Also it must be stressed that in the presentation so far the discrete systems have been written so that they express quantities *per unit volume*. As has already been pointed out, finite volume methods usually construct discrete systems which approximate the total fluxes and forces on each CV. Therefore after restriction of the velocity and pressure obtained on the fine grid and application of the coarse Navier-Stokes operator one obtains the product $\Delta\Omega_P \cdot \tau_{2h}{}^h{}_{,P}$ for each coarse CV. This quantity must be divided by the volume $\Delta\Omega_P$ of each CV to obtain $\tau_{2h}{}^h$ before (3.12) can be applied.

Finally, it is appropriate to discuss the implications of the approximate solution of the discrete systems by iterative solvers. Indeed, it is not possible in general to solve the discrete Navier-Stokes system exactly, but the residual may be made as small as desired, up to machine precision, by performing an appropriate number of iterations. If $\phi_h^{*k}$ is the approximate solution to system (2.3) after iteration $k$ and $r_h^k$ is the associated residual then:

$$N_h \phi_h^{*k} = b_h - r_h^k \tag{3.13}$$

Subtracting (3.13) from (2.2), and (2.3) from (2.2) one gets respectively:

$$N_h \phi_h - N_h \phi_h^{*k} = -\tau_h + r_h^k \tag{3.14}$$

$$N_h \phi_h - N_h \phi_h^* = -\tau_h \tag{3.15}$$

Comparing (3.14) with (3.15) it is easy to see that the solution $\phi_h^{*k}$ corresponds to a 'truncation error' $\tau_h - r_h^k$, just as solution $\phi_h^*$ corresponds to the truncation error $\tau_h$. To attain the full accuracy that a finite volume method can offer the residual should be reduced to the level of the truncation error in every CV of the grid. Furthermore to accurately estimate the truncation error, the residual should be smaller than the truncation error in every CV, say $r_h^k{}_{,P} \leq 0.1 \cdot \tau_{h,P}$ for every CV $P$. Therefore, for convergence of the iterative method one should not monitor the mean residual but the residual / truncation error ratio in every CV. The residual acts as a source of algebraic error $\phi_h^* - \phi_h^{*k}$, just as the truncation error acts as a source of discretisation error – see [12]. Therefore, a high residual in one region may generate a high algebraic error in another where the residual itself is small. On the other hand there is no point in reducing the residual far below the truncation error, as (3.14) and (3.15) indicate: The algebraic error $\phi_h^* - \phi_h^{*k}$ would reduce, but the exact error $\phi_h - \phi_h^{*k}$ would not. Again, usually the discrete systems of finite volume methods are such that the quantity $\Delta\Omega_P \cdot r_{h,P}^k$ is more easily attainable for each CV.

## 4. 2$^{nd}$-order finite volume discretisation for the Navier – Stokes equations

Here the particular discretisation schemes which will be used in section 6 to test the method are briefly described. The 2-D stationary incompressible Navier-Stokes equations under constant density $\rho$ and viscosity $\mu$, integrated over a CV $P$ of volume $\Delta\Omega_P$ are written in cartesian coordinates as:

$$N_{h,P}^x(u,v,p) \equiv \frac{1}{\Delta\Omega_P}\left[\oint_{S_P}\rho\mathbf{V}\cdot\mathbf{n}\,u\,dS - \oint_{S_P}\mu\nabla u\cdot\mathbf{n}\,dS + \oint_{S_P}p\mathbf{i}\cdot\mathbf{n}\,dS\right] = 0 \qquad (4.1)$$

$$N_{h,P}^y(u,v,p) \equiv \frac{1}{\Delta\Omega_P}\left[\oint_{S_P}\rho\mathbf{V}\cdot\mathbf{n}\,v\,dS - \oint_{S_P}\mu\nabla v\cdot\mathbf{n}\,dS + \oint_{S_P}p\mathbf{j}\cdot\mathbf{n}\,dS\right] = 0 \qquad (4.2)$$

$$N_{h,P}^m(u,v) \equiv \frac{1}{\Delta\Omega_P}\oint_{S_P}\rho\mathbf{V}\cdot\mathbf{n}\,dS = 0 \qquad (4.3)$$

where $S_P$ is the surface of CV $P$, $\mathbf{n}$ is the outward normal unit vector at each point of the surface, $\mathbf{i}$ and $\mathbf{j}$ are the unit vectors in the $x$- and $y$- directions, $u$ and $v$ are the components of the velocity vector $\mathbf{V} = u\mathbf{i} + v\mathbf{j}$, and $p$ is the pressure.

The boundary of each CV will be composed of a number of straight faces, each of which separates it from another single CV or from the exterior of the computational domain. Figure 1 shows a face $f$ separating two CVs, with centroids $P$ and $N$. The centre of the face is denoted by $c$, and $c'$ denotes the point on the line $PN$ which is closest to $c$. Also points $P'$ and $N'$ are such that the segment $P'N'$ is of the same length as $PN$, and is perpendicular to the face $f$, and its midpoint is point $c$. The part of the grid shown in figure 1 exhibits *skewness*, that is the line joining $P$ and $N$ does not pass through the centre $c$ of face $f$. It is also non-orthogonal, which means that the angle $\theta$ between $PN$ and the face normal is non-zero. Finally, if the middle of the line segment $PN$ is far from face $f$ then the grid will also be said to exhibit *expansion*.

The gradient operator is frequently used in discretisation schemes and here it will be approximated using the least squares method suggested in [4]. This method assigns to the discrete gradient $\nabla_h$ of the variable $\phi_h$ at the centre of a CV $P$ the appropriate value so that the sum $\Sigma_N\{[\Delta\phi_N - (\nabla_h\phi_h)_P\cdot\Delta\mathbf{r}_N] / |\Delta\mathbf{r}_N|\}^2$ is minimised (the index $N$ runs through all neighbours of $P$, and $\Delta\phi_N = \phi_{h,N} - \phi_{h,P}$, $\Delta\mathbf{r}_N = N - P$). See [11] for an explicit expression for $\nabla_h$ in the two-dimensional case. In the following, $\nabla_h^x$, $\nabla_h^y$ will denote the two cartesian components of $\nabla_h$.

The Navier-Stokes equations will be discretised by approximating the fluxes and forces on each CV face, using the same or similar schemes as in [12]. In the following, an overbar denotes a value obtained by linear interpolation at point $c'$ from the values at points $P$ and $N$, and a subscript $c$ denotes a kind of linear interpolation, suggested in [12], which accounts for skewness and approximates the value at point $c$ as:

$$\phi_{h,c} = \overline{(\phi_h)}_{c'} + \overline{(\nabla_h\phi_h)}_{c'}\cdot(\mathbf{c}-\mathbf{c'}) \qquad (4.4)$$

Also, the value of $\phi$ at point $P'$ is approximated as:

$$\phi_{h,P'} = \phi_{h,P} + (\nabla_h \phi_h)_P \cdot (\boldsymbol{P'} - \boldsymbol{P}) \qquad (4.5)$$

Then, the various terms of the *x*-momentum equation are discretised as:

$$\frac{1}{\Delta\Omega_P} \oint_{S_P} \rho \boldsymbol{V} \cdot \boldsymbol{n} u \, dS \approx \frac{1}{\Delta\Omega_P} \sum_{f \in f_P} F_{h,f} \cdot u_{h,c} \equiv C_{h,P}^{x*}(u_h, v_h, p_h) \qquad (4.6)$$

$$\frac{1}{\Delta\Omega_P} \oint_{S_P} \mu \nabla u \cdot \boldsymbol{n} \, dS \approx \frac{1}{\Delta\Omega_P} \sum_{f \in f_P} \mu S_f \frac{u_{h,N'} - u_{h,P'}}{|\boldsymbol{N'} - \boldsymbol{P'}|} \equiv D_{h,P}^{x*}(u_h) \qquad (4.7)$$

$$-\frac{1}{\Delta\Omega_P} \oint_{S_P} p \boldsymbol{i} \cdot \boldsymbol{n} \, dS \approx -\frac{1}{\Delta\Omega_P} \sum_{f \in f_P} p_{h,c} n_f^x S_f \equiv P_{h,P}^{x*}(p_h) \qquad (4.8)$$

In the above, $f_P$ is the set of all faces of CV $P$ and $n_f^x$, $n_f^y$ are the cartesian components of the outward unit vector $\boldsymbol{n}_f$ which is perpendicular to $f$. Also $F_{h,f}$ is the discrete mass flux through face $f$, to be defined shortly. The *y*-momentum equation is discretised analogously, while the continuity equation is discretised as:

$$\frac{1}{\Delta\Omega_P} \oint_{S_P} \rho \boldsymbol{V} \cdot \boldsymbol{n} \, dS \approx \frac{1}{\Delta\Omega_P} \sum_{f \in f_P} F_{h,f} \equiv N_{h,P}^{m*}(u_h, v_h, p_h) \qquad (4.9)$$

The sum of the approximate discrete operators (4.6)-(4.8) is the discrete *x*-momentum operator $N_h^{x*}(u_h,v_h,p_h) = C_h^{x*}(u_h,v_h,p_h) - D_h^{x*}(u_h) - P_h^{x*}(p_h)$, which tries to approximate the exact *x*-momentum operator $N_h^x$ (4.1). The associated truncation error with respect to the exact solution $(u,v,p)$ of (4.1)-(4.3) is $\tau_h^x = N_h^x(u,v,p) - N_h^{x*}(u_h,v_h,p_h)$. Similarly, the discrete *y*-momentum and continuity operators approximate the exact operators (4.2) and (4.3) up to truncation errors $\tau_h^y = N_h^y(u,v,p) - N_h^{y*}(u_h,v_h,p_h)$, $\tau_h^m = N_h^m(u,v) - N_h^{m*}(u_h,v_h,p_h)$. The truncation errors $\tau_h^x$, $\tau_h^y$, $\tau_h^m$ will be estimated using (3.12). The above schemes are in general considered to have truncation errors of $O(h^2)$ – see [12] – so $p = 2$ will be used in (3.12).

Most of the above discretisation schemes use linear interpolation, which has the effect that the image $L_h \phi_h$ of a discrete operator $L_h$ which uses it may be smooth even if $\phi_h$ contains a component which oscillates from CV to CV (i.e. with period of oscillation equal to two CVs). Or equivalently, the solution $\phi_h$ of the system $L_h \phi_h = b_h$ may contain oscillations even if $b_h$ is smooth. A special case of this is the so-called checkerboard distribution depicted in figure 2: If, for example, the pressure at the CV centres assumes the values shown in the figure, then obviously linear interpolation gives zero pressure at the face centres, and the operator $P_h^{x*}$ (4.8) gives zero pressure force on each CV. At domain boundaries pressure is extrapolated from the interior, and such an oscillating pressure field would result in non-zero oscillating forces along the boundary CVs, which means that an oscillating pressure field is not part of the null space of $P_h^{x*}$. Therefore one may be tempted to think that according to (3.15), as $h \to 0$ if $\tau_h \to 0$ then $p_h^*$ will tend to the exact pressure $p_h$ which is oscillations-free. However, the oscillating pressure field is close to being an eigenvector of $P_h^{x*}$ corresponding to a zero eigenvalue, and the smaller the grid spacing $h$ the closer it is to such an eigenvector. In practice this means that pressure oscillations may indeed appear in the discrete solution and they may be very resistant to grid refinement.

A similar, but not as bad, situation holds also for the velocity field. The convection operators (4.6) and (4.9) produce images which may be smooth even if the velocity components oscillate at CV centres. However the discrete viscous force operator (4.7) involves direct velocity differences between adjacent CV centres and therefore always reflects velocity oscillations to its image. Consequently, the phenomenon of oscillations in the $u_h^*$, $v_h^*$ fields becomes less intense as the Reynolds number decreases, and in fact oscillations diminish with grid refinement, and are rarely a problem for incompressible flows. The discrete gradient operator $\nabla_h$ may also produce a smooth image when applied to an oscillating field.

Pressure oscillations are a serious problem for colocated grids which has been addressed by many researchers. One way around it is to observe that the pressure forces are calculated from values of pressure estimated at face centres using linear interpolation. These values have a much smaller discretisation error than the oscillating values at CV centres. So after obtaining the solution to the discrete system one may discard the pressures at the CV centres and consider the pressure field to be given by the pressures at the face centres. But rather than obtaining an oscillating pressure field and eliminating the oscillations afterwards it is more desirable to obtain an oscillations-free field altogether, to avoid problems for the algebraic solvers of the discrete system. One possibility is to use another discrete operator for the pressure force, one which reflects pressure oscillations to its image, like the one proposed in [13]. However, the most popular method involves adding an artificial pressure term to the discrete expression for the mass flux through a face, generally known as momentum interpolation.

Momentum interpolation was originally proposed in [14]. Since then many variants of this technique have been proposed but most of them share the feature that the discretisation of the face mass fluxes is interlinked with a SIMPLE-like solution method (there are a few exceptions, e.g. [10]). SIMPLE-like solution algorithms linearise the momentum equations to obtain linear systems for the velocity components, whose $P$-th equation has the form:

$$A^u_{P,P} u_{h,P} + \sum_N A^u_{P,N} u_{h,N} = Q^{u,\backslash p}_P - \sum_{f \in \mathrm{f}_P} p_{h,c} S_f \, \boldsymbol{n}_f \cdot \boldsymbol{i} \tag{4.10}$$

where $N$ runs over all neighbours of CV $P$, and $A^u_{ij}$ is the $(i,j)$-th coefficient of the matrix of coefficients of the linear system for $u$. If face $f$ separates CVs $P$ and $N$ and $\boldsymbol{n}_f$ points from $P$ to $N$ (see figure 3) then the momentum interpolation variant of [12], which is more appropriate for our discretisation, approximates the mass flux through $f$ as:

$$F_{h,f} = \rho_c S_f \cdot \left( \boldsymbol{V}_{h,c} \cdot \boldsymbol{n}_f + a_{mi} \frac{S_f}{A^u_f} \left[ (p_{h,P} - p_{h,N}) - \frac{1}{2} \left[ (\nabla_h p_h)_P + (\nabla_h p_h)_N \right] \cdot (\boldsymbol{P} - \boldsymbol{N}) \right] \right) \tag{4.11}$$

where:

$$A^u_f = \frac{1}{2} \left( A^u_{P,P} + A^u_{N,N} \right) \tag{4.12}$$

Here $\boldsymbol{V}_{h,c} = u_{h,c} \boldsymbol{i} + v_{h,c} \boldsymbol{j}$, and $a_{mi}$ is a real factor introduced for better control of the pressure term. Most researchers use $a_{mi} = 1$. Obviously (4.11) is equivalent to interpolation (4.4) but with the addition of a pressure term. However, mass fluxes are functions of velocity only and therefore the pressure term relates completely to the truncation error. Therefore the magnitude of the pressure term should diminish as $h \to 0$, at a rate which is at least $2^{\text{nd}}$-order to preserve the overall order of accuracy of the discretisation. Indeed using Taylor series one can show that, if $\nabla_h$ is at least $2^{\text{nd}}$-order accurate:

$$(p_{h,P} - p_{h,N}) = \frac{1}{2} \left[ (\nabla_h p_h)_P + (\nabla_h p_h)_N \right] \cdot (\boldsymbol{P} - \boldsymbol{N}) + O(h^3) \tag{4.13}$$

In addition $S_f \in O(h)$ and $A^u_f \in O(1)$, so the pressure-part of the discrete mass flux is:

$$a_{mi} \rho_c \frac{S_f^2}{A^u_f} \left[ (p_{h,P} - p_{h,N}) - \frac{1}{2} \left[ (\nabla_h p_h)_P + (\nabla_h p_h)_N \right] \cdot (\boldsymbol{P} - \boldsymbol{N}) \right] \in O(h^5) \tag{4.14}$$

By dividing by the volume $\Delta\Omega \in O(h^2)$ of a CV which shares the face $f$ one sees that (4.14) contributes a $O(h^3)$ component to the truncation error. In fact, in [11] it is shown that under special but not uncommon circumstances the sum of the terms (4.14) for two opposite faces of a quadrilateral CV becomes $O(h^6)$ because the leading terms of their Taylor expansions cancel out. This corresponds to a $O(h^4)$ contribution to the truncation error.

The pressure-part (4.14) of the discrete mass flux consists of the difference between two parts, one involving the direct pressure difference $p_{h,P} - p_{h,N}$ between the centroids of the adjacent CVs, and one involving the pressure gradient. The part involving the pressure gradient is again insensitive to pressure oscillations, but the part involving the direct pressure difference is not. Indeed, if pressure oscillates from one CV to the other then the pressure difference will also oscillate from face to face, and so will the discrete mass flux. Therefore, the discrete Navier-Stokes and continuity operators do reflect pressure oscillations to their image, which removes pressure oscillations from the discrete solution.

The main disadvantage of expression (4.11) is that to calculate the mass fluxes $F_h$ one needs the coefficients of the matrix $A^u$, but to calculate the coefficients of $A^u$ one needs the mass fluxes $F_h$! Therefore, given a discrete flow field $u_h$, $v_h$, $p_h$ one cannot directly evaluate the mass fluxes through the faces of the CVs but has to resort to an iterative procedure. For our truncation error estimator this means that the expression (4.14) cannot readily be evaluated on the coarse grid $2h$. One may argue that since the magnitude of the pressure term of $F_h$ reduces at a rate which is faster than 2nd-order then it may simply be omitted on grid $2h$. On the other hand, including this term on grid $2h$ allows for a cleaner approach which may also be used with up to 4th-order accurate overall discretisation schemes. Indeed, schemes based on higher-order rather than linear interpolation may also allow for oscillating pressure fields, and this is why in [15] momentum interpolation is used in the context of a 4th-order accurate discretisation.

## 5. New momentum interpolation

One idea to overcome the above problem is to use in (4.11) only the viscous-part of $A^u$ which contains only geometric terms and does not depend on $F_h$. However, this leads to the coefficient of the pressure-term of the mass flux being too big, resulting in divergence of the solution method unless a very small value of $a_{mi}$ is used (in [16] $a_{mi} = 0.04$ is used). This in turn was found not to eliminate the pressure oscillations at some regions of the flow field. Therefore, the velocity field has to be taken into account, and this is done through $A^u$. The reasoning behind the choice of the coefficient of the pressure term is the following: Equation (4.10) suggests that the contribution of pressure to the value of $u$ at point **P** is:

$$\delta^p u_{h,P} = -\frac{1}{A^u_{P,P}} \sum_{f \in f_P} p_{h,c} S_f \, \boldsymbol{n}_f \cdot \boldsymbol{i} \approx -\frac{1}{A^u_{P,P}} \left(\nabla^x_h p_h\right)_P \cdot \Delta\Omega_P \qquad (5.1)$$

where $\delta^p u_{h,P}$ is the part of $u_{h,P}$ which is due to pressure forces. The (discrete) Gauss theorem is used to obtain the second equality. Of course the above assumption is very crude because in (4.10) the coefficients of $A^u$ are also functions of $u_h$ and $v_h$, and $u_{h,N}$ also depend on the pressure field, and also upwind differencing is used to form $A^u$ while the central difference scheme is imposed through deferred correction. However (5.1) gives a feel of the importance of pressure in determining $u_h$. A similar relation can be derived for $v_h$, and in fact the coefficients of $A^v$ are nearly equal to the coefficients of $A^u$, except maybe near some boundaries. Therefore, in [12] the assumption is made that a similar relation holds for the component of velocity in any direction. In particular if $u^n_{h,c}$ is the component of velocity normal to face $f$ at $c$, $u^n_{h,c} = \boldsymbol{V}_{h,c} \cdot \boldsymbol{n}_f$, then it is assumed that:

$$\delta^P u_{h,c}^n = -\frac{1}{A_f^u} (\nabla_h p_h)_c \cdot \boldsymbol{n}_f \cdot \Delta\Omega_f \qquad (5.2)$$

where the product $(\nabla_h p_h)_c \cdot \boldsymbol{n}_f$ equals the pressure gradient in the direction of $\boldsymbol{n}_f$ at the face centre. The volume $\Delta\Omega_f = S_f(\boldsymbol{N}-\boldsymbol{P})\cdot\boldsymbol{n}_f$ is defined as the volume of the imaginary CV around face $f$ depicted by dashed line in figure 3, which has two sides parallel to face $f$ and passing through points $\boldsymbol{P}$ and $\boldsymbol{N}$, and two sides perpendicular to $f$ passing through its vertices. If one further approximates $(\nabla_h p_h)_c$ as the mean of $(\nabla_h p_h)_P$ and $(\nabla_h p_h)_N$ then (5.2) becomes:

$$\delta^P u_{h,c}^n = \frac{S_f}{A_f^u}\left[\frac{1}{2}\left[(\nabla_h p_h)_P + (\nabla_h p_h)_N\right]\cdot(\boldsymbol{P}-\boldsymbol{N})\right] \qquad (5.3)$$

Actually $\boldsymbol{P'}$ and $\boldsymbol{N'}$ (see figure 3) should be used instead of $\boldsymbol{P}$ and $\boldsymbol{N}$ in (5.3) but since the pressure contribution is very approximate this substitution is acceptable. Because of (4.13) one can substitute (5.3) by:

$$\delta^P u_{h,c}^n = \frac{S_f}{A_f^u}(p_{h,P} - p_{h,N}) \qquad (5.4)$$

By substituting the pressure contribution (5.3) to the normal component of velocity by (5.4) and using the midpoint rule that $F_{h,f} = \rho_c S_f u_{h,c}^n$ one arrives at (4.11).

The above reasoning, although it does not sound very solid, in practice gives an appropriate magnitude to the coefficient of the pressure term of the discrete mass fluxes, with $a_{mi} \approx 1$. If $a_{mi}$ is much smaller then the method fails to eliminate the oscillations, while if $a_{mi}$ is much larger then the system is difficult to solve and divergence occurs on grids of reasonable fineness (of course if the grid is fine enough then the significance of the pressure term will diminish no matter what the value of $a_{mi}$).

Now, to uncouple the mass flux discretisation from the iterative solution method, in the present work $A_f^u$ will be substituted in (4.11) by a pseudo-coefficient $A_f$ which depends directly on the grid geometry around face $f$ and on the velocity at the adjacent CVs. Just as $A_{P,P}^u$ is the coefficient by which $u_{h,P}$ is multiplied in the linearised discrete $x$-momentum equation of CV $P$, $A_f$ is constructed as the coefficient by which $u_{h,c}^n$ would be multiplied in a hypothetical linearised discrete $n$-momentum equation for the imaginary CV around $f$ in figure 3. The construction of the hypothetical momentum equation proceeds as follows. First define the points $\boldsymbol{PP'}$, $\boldsymbol{NN'}$, $\boldsymbol{VV1}$, $\boldsymbol{VV2}$ such that $\boldsymbol{P'}$ lies midway between $c$ and $\boldsymbol{PP'}$ etc. (see figure 3). Starting with viscous forces, for the side of the imaginary CV which passes through $N$, the $n$-component is discretised as:

$$\int_N \mu \nabla u^n \cdot \boldsymbol{n}\, dS \approx \mu \frac{u_{h,NN'}^n - u_{h,c}^n}{|NN'-c|} S_f = \mu \frac{u_{h,NN'}^n - u_{h,c}^n}{2|N'-c|} S_f \qquad (5.5)$$

where $u^n$ is the component of velocity normal to face $f$. A similar scheme will be used for the viscous component over the face through $\boldsymbol{P}$. For the face through $V1$ it will be assumed that:

$$\int_{V1} \mu \nabla u^n \cdot \boldsymbol{n}\, dS \approx \mu \frac{u_{h,VV1}^n - u_{h,c}^n}{2|V1-c|} S_V \qquad (5.6)$$

where $S_V = (N–P) \cdot n_f$ is the length of each of the faces which are perpendicular to $f$. A similar assumption is made for the face which passes through $V2$. Therefore the total contribution of viscous forces to the coefficient of $u^n_{h,c}$ is:

$$A_f^{visc} = \frac{\mu \cdot S_f}{2|N'-c|} + \frac{\mu \cdot S_f}{2|P'-c|} + \frac{\mu \cdot S_V}{2|V1-c|} + \frac{\mu \cdot S_V}{2|V2-c|} \qquad (5.7)$$

This can be simplified because $2|V1-c| = 2|V2-c| = S_f$. Also since one is only interested in an approximate value for $A_f$, (5.7) can be further simplified by assuming that $c$ lies midway between $N'$ and $P'$ so that $2|N'-c| = 2|P'-c| = (N-P) \cdot n_f = S_V$. Therefore (5.7) becomes:

$$A_f^{visc} = 2\mu \cdot \left[ \frac{S_f}{S_V} + \frac{S_V}{S_f} \right] \qquad (5.8)$$

If $\rho$ and/or $\mu$ are not constant, then the Navier-Stokes equations include other viscous force components as well, but usually in the SIMPLE framework they do not contribute to the matrices $A^u$, $A^v$. Therefore the viscous contribution (5.8) remains the same.

For convection, in the original SIMPLE method the coefficients of $A^u$ are formed using the upwind difference scheme while the central difference scheme is enforced through deferred correction. Therefore, an upwind-like approach will be used for the convective part of $A_f$. It is assumed that the velocity at the centre of each face of the imaginary CV equals $V_{h,c}$ and that the mass flux through each face is given by the usual midpoint rule, $F = V_{h,c} \cdot n \cdot S$. Whatever the direction of $V_{h,c}$ mass will flow out of the imaginary CV only through two of the faces, one parallel to $f$ and one perpendicular. Therefore, according to the upwind philosophy only these two faces will contribute to $A_f$. If by rotating the vector $x$ 90° anticlockwise one gets the vector $rot(x)$, then $rot(n_f)$ is the unit vector which is perpendicular to the faces which pass through $V1$ and $V2$, and the sum of the convective contributions to $A_f$ is:

$$A_f^{convec} = \rho_c S_f \left| V_{h,c} \cdot n_f \right| + \rho_c S_V \left| V_{h,c} \cdot rot(n_f) \right| \qquad (5.9)$$

It is reminded that $S_V = (N-P) \cdot n_f$. In total the value of $A_f$ is therefore:

$$A_f = A_f^{convec} + A_f^{visc} \qquad (5.10)$$

The proposed momentum interpolation is therefore to use (4.11) with $A^u_f$ replaced by $A_f$ given by (5.10), (5.9) and (5.8). The new method also overcomes the well-known problem of the dependency of (4.11) on the underrelaxation factor of the solution method and the time step for transient flows – see [17], [18].

Before ending this section it should be mentioned that at first there was an effort to use central differencing for the convective part of $A_f$ which resulted in:

$$A_f^{convec} = \frac{1}{2} (V_{h,N} - V_{h,P}) \cdot n_f \, S_f \qquad (5.11)$$

However, this did not work. It seems that (5.11) gives too small a value for the convective part because for smoothly varying fields the difference $V_{h,N} - V_{h,P}$ will be small. Therefore, $A_f$ will again be dominated by the viscous, geometric terms and the problems mentioned earlier will arise. This also highlights the importance of using upwind differencing for convection for the construction of the matrix of coefficients of several solution methods including SIMPLE.

# 6. Testing of the method

The method is tested on two cases with analytic solution, which allows direct comparison between the truncation error estimate and the actual truncation error: A particular lid-driven cavity problem, and the flow between concentric cylinders.

## *6.1 Lid-driven cavity*

*6.1.1 Momentum interpolation*

We start with a few comments on the new variant of momentum interpolation. Extensive results will not be presented because it was observed that in general this variant offers nearly identical accuracy and rates of algebraic convergence (if SIMPLE is used) as classic momentum interpolation. Here are presented some results of applying the method to the simulation of the flow in a skew lid-driven cavity of side L = 1 m with side walls inclined at 45°, and the top lid moving at $V_{lid}$=1 m/s (see [19]), at Re=1000 ($\rho$=1 kg/m$^3$, $\mu$=0.001 Pa·s). The pressure is fixed to zero at $(x,y) = (0.5, 0.01)$, where the origin $(x,y) = (0,0)$ is at the lower-left corner. This is not the lid-driven cavity problem with analytic solution which will be used in the next section to assess the truncation error estimator, but it allows testing of the momentum interpolation on grids which are not cartesian.

The problem was solved using uniform grids, with three different schemes for the mass fluxes: 1) linear interpolation (mi0) – i.e. with term (4.14) completely absent from (4.11), 2) classic momentum interpolation (mi1), and 3) new momentum interpolation (mi2). Figure 4 shows the pressure distribution along the horizontal line passing through the centres of the CVs which lie immediately above the line $y$=H·3/4, where H = L·sin45° is the height of the cavity, on two grids of different density.

If mi0 is used then pressure oscillations appear in the solution, whose amplitude increases towards the interior of the domain, and which do not diminish with grid refinement. In this case the oscillating component of the pressure field is close to being an eigenvector corresponding to the zero eigenvalue, which means that small changes in the algebraic residual may reflect large changes in the amplitude of the oscillations. To minimise the possibility that the oscillations are a product of insufficient residual reduction, iterations were continued until the residual was below $10^{-5}$ N/m$^3$ in each CV of the grid. Also the solution was obtained using two different initial estimates, one being the prolonged solution of the immediately coarser grid, and the other being the smooth solution obtained with momentum interpolation, but no significant differences were observed. If momentum interpolation is used then a smooth pressure field is obtained, and on the grid 128×128 the results of the two variants of momentum interpolation are indistinguishable.

Pressure oscillations also have a detrimental impact on the convergence rate of the SIMPLE algorithm, which was used to solve the discrete systems. For example, to solve the problem on the 64×64 grid up to the $10^{-5}$ residual, using the solution of the 32×32 grid as initial estimate, 24000 iterations were required in the "mi0" case, as opposed to 485 and 479 iterations in cases "mi1" and "mi2" respectively (underrelaxation factors $a_u$=0.8, $a_p$=0.3, and a 2$^{nd}$ pressure correction for grid non-orthogonality were used as suggested in [12]). If multigrid is used things would be even worse as pressure oscillations which developed in coarse grids would be prolonged to the fine ones. Also it is very important to note that for the coefficients of the pressure-correction system of the SIMPLE algorithm, (4.12) must be used instead of (5.10), otherwise the method diverges. This may sound strange since both formulae give similar values at convergence, but they may differ significantly at the first stages of iteration because (4.12) is computed from mass fluxes which are calculated after the pressure correction step, while (5.10) from velocities obtained before this step. Finally, it was found that with "mi2" for extremely coarse grids it may be necessary to use $a_{mi} < 1$ otherwise SIMPLE may diverge, unlike if "mi1" is used, because $A_f$ (5.10) is a little smaller than $A^u_f$ (4.12) (due to the fact that the velocity underrelaxation factor is not taken into account in $A_f$, unlike $A^u$). This holds also for multigrid methods which use very coarse grids.

*6.1.2 Analytic solution*

To test the truncation error estimator we apply it to the lid-driven cavity problem of [20], which has an analytic solution: Fluid of $\rho=1$ kg/m$^3$, $\mu=0.001$ Pa·s is enclosed in a square cavity, whose sides of length L=1 m are aligned with the $x$- and $y$- axes. The top wall (lid) moves with a horizontal velocity $u(x,1) = 16(x^4–2x^3+x^2)$, and there exists a body force $b$ in the $y$-direction:

$$b(x,y) = 8\mu[24F(x) + 2f'(x)g''(y) + f'''(x)g(y)] + 64[F_2(x)G_1(y) - g(y)g'(y)F_1(x)] \quad (6.1)$$

where:

$$f(x) = x^4 - 2x^3 + x^2$$
$$g(y) = y^4 - y^2$$
$$F(x) = \int f(x)dx$$
$$F_1(x) = f(x)f''(x) - [f'(x)]^2$$
$$F_2(x) = \int f(x)f'(x)dx = 0.5[f(x)]^2$$
$$G_1(y) = g(y)g'''(y) - g'(y)g''(y)$$

where the primes denote differentiation. The exact solution to this problem (eqns. (4.1)-(4.3) with the addition of $(-\int_{\Delta\Omega_P}bd\Omega)/\Delta\Omega_P$ in the left-hand side of (4.2)) is:

$$u(x,y) = 8f(x)g'(y) \quad (6.2)$$
$$v(x,y) = -8f'(x)g(y) \quad (6.3)$$
$$p(x,y) = 8\mu[F(x)g'''(y) + f'(x)g'(y)] + 64F_2(x)\{g(y)g''(y) - [g'(y)]^2\} \quad (6.4)$$

(actually any pressure field $p' = p + c$ will do, for any constant c). The problem is discretised using the schemes of sections 4-5, plus the body force is discretised with the midpoint rule as $\int_{\Delta\Omega_P}bd\Omega \approx b(x_P,y_P)\cdot\Delta\Omega_P$. Since the exact solution is known, the exact truncation errors $\tau_h^x$, $\tau_h^y$, $\tau_h^m$ of the operators $N_h^{x*}$, $N_h^{y*}$, $N_h^{m*}$ can be calculated. The calculation of $\tau_h^y$ requires integration of the body force over each CV, so we will focus on $\tau_h^x$, $\tau_h^m$.

The problem is solved on a uniform and a non-uniform series of structured grids of up to 256×256 CVs. The non-uniform grids are such that, if each CV is assigned a horizontal index $i$ and a vertical index $j$, and $\Delta x_h^{i,j}$, $\Delta y_h^{i,j}$ are the horizontal and vertical sizes of CV $(i,j)$ of grid $h$ respectively, then there is a constant $r_h$ such that $\Delta x_h^{i+1,j} / \Delta x_h^{i,j} = r_h$ for $x < 0.5$ and $\Delta x_h^{i+1,j} / \Delta x_h^{i,j} = 1/r_h$ for $x > 0.5$, and similarly for $\Delta y$. The expansion ratio $r_h$ for grid 256×256 is such that the boundary CVs which touch the centrelines have a ratio of $\Delta x/\Delta y = 10:1$ or 1:10. Grid $2h$ comes from grid $h$ by removing every second grid line, so for the non-uniform grids $r_h = \sqrt{r_{2h}}$ and $r_h$ ranges from about 1.156 on grid 32×32 (this grid is shown in figure 5) to about 1.018 on grid 256×256. Algebraic residuals were dropped below $10^{-8}$ in every CV. Figure 6 shows the exact distributions of $\tau_h^x$ and $\tau_h^m$ on the 256×256 grids.

In the following, $\tau_h^{x*}$, $\tau_h^{y*}$, $\tau_h^{m*}$ will denote the truncation error estimates (3.12), and $\varepsilon_h^u$, $\varepsilon_h^v$, $\varepsilon_h^p$ the discretisation errors. Figures 7 and 8 (left) show the distributions of $\varepsilon_h^u$ along the centres of the CVs which lie just to the right of the vertical centreline ($x=0.5$), and the distributions of $\varepsilon_h^p$ ($\varepsilon_h^p$ was set equal to zero at the centre of the CV of the lower left corner of the domain) along the centres of the CVs which lie just above the horizontal centreline ($y=0.5$). The errors are displayed in logarithmic scale, and from the distance between the distributions it is verified that the particular finite volume method is 2$^{nd}$-order accurate. Convergence is not as regular for the non-uniform grids as for the uniform ones, and in fact the discretisation errors on coarse non-uniform grids decrease at a rate which is faster than 2$^{nd}$-order, but tends to become 2$^{nd}$-order with refinement.

For the calculation of $\tau_{2h}^h$ (3.5) two different restriction operators were used. The first is proposed in [12]:

$$\left(I_h^{2h}\phi_h^*\right)_P = \frac{1}{4}\sum_{C\in C_P}\left(\phi_{h,C}^* + \left(\nabla_h\phi_h^*\right)_C\cdot(\boldsymbol{P}-\boldsymbol{C})\right) \tag{6.5}$$

where $C_P$ is the set of 4 CVs of grid $h$ which cover CV $P$ of grid $2h$ - the CVs of the set $C_P$ will henceforth be called the *children* of the *parent P*. The other is proposed in [11]:

$$\left(I_h^{2h}\phi_h^*\right)_P = \frac{1}{4}\sum_{C\in C_P}\left\{\phi_{h,C}^* + \left(\nabla_h\phi_h^*\right)_C\cdot(\boldsymbol{P}-\boldsymbol{C}) + \frac{1}{2}\left[\left(\phi_{xx}^*\right)_{h,C}\cdot\Delta x_C^2 + \left(\phi_{yy}^*\right)_{h,C}\cdot\Delta y_C^2 + \left(\left(\phi_{xy}^*\right)_{h,C} + \left(\phi_{yx}^*\right)_{h,C}\right)\cdot\Delta x_C\cdot\Delta y_C\right]\right\} \tag{6.6}$$

where $\Delta x_C = x_P - x_C$ etc., and $(\phi_{xx}^*)_h$ etc. are the approximate second derivatives of $\phi^*$ on grid $h$, which are calculated by applying a least squares differentiation again to the components of $\nabla_h\phi_h^*$. For the second differentiation, the neighbours of each CV $C$ are considered to be its *siblings* (i.e. the CVs which have the same parent as $C$). In [11] it is shown that (6.5) is 2nd-order accurate as long as $\nabla_h$ is at least 1st-order accurate, while (6.6) is 2nd-order accurate if $\nabla_h$ is 1st-order, and 3rd-order if $\nabla_h$ is at least 2nd-order. In our case $\nabla_h$ is 2nd-order accurate except for the boundary CVs where it is 1st-order. For prolongation in (3.12), the following 2nd-order accurate operator was used:

$$\left(I_{2h}^h\phi_{2h}\right)_C = \phi_{2h,P} + \left(\nabla_{2h}\phi_{2h}\right)_P\cdot(\boldsymbol{C}-\boldsymbol{P}) \tag{6.7}$$

The vertical centreline is not appropriate for the study of the truncation error because certain derivatives of the flow variables become zero there and this causes the leading term of the truncation error to diminish, as shown in figure 6. This is shown clearly in figure 9, which shows $|\tau_h^m|$ along the centres of the CVs just to the right of the vertical centreline, for a series of uniform grids: the distance between the distributions of consecutive grids indicates that $\tau_h^m$ reduces almost at a 4th-order rate. The estimate $|\tau_h^{m*}|$ is also plotted in the same figure, and it is clear that it captures the overall shape of $\tau_h^m$ but it is always about 4 times higher. This is not surprising since the assumption that $p=2$ is made in (3.12).

In figure 10, $|\tau_h^x|$ is displayed along the CV centres just to the right of $x=0.75$ on the uniform grids. Here $\tau_h^x$ indeed reduces at a 2nd-order rate. Two estimates $\tau_h^{x*}$ are also shown, one using (6.5) and one using (6.6). The one using (6.6) appears to be more accurate, which is verified by the graph of the quantity $|(\tau_h^x-\tau_h^{x*})/\tau_h^x|$. Since the estimate (3.12) is based on the assumption that the magnitude of the truncation error is determined by its leading term, the difference $\tau_h^x-\tau_h^{x*}$ should be of the order of the second leading term of the truncation error, and therefore $|(\tau_h^x-\tau_h^{x*})/\tau_h^x|$ should be $O(h)$ (except if only odd or even powers of $h$ appear in the expansion of $\tau_h^x$, in which case $|(\tau_h^x-\tau_h^{x*})/\tau_h^x|$ should be $O(h^2)$). For the estimate using (6.5) this quantity reduces at a rate which is less than 1st order while for the estimate using (6.6) it reduces at a rate which is faster than 1st order (but less than 2nd order). At $y\approx 0.32$ and $y\approx 0.8$ this quantity does not reduce, because the truncation error there reduces at a rate which is faster than 2nd-order (actually it changes sign – see figure 6). Therefore, a situation similar to that shown in figure 9 occurs there. Similar conclusions are drawn from figure 10 concerning the distributions of $\tau_h^m$ and $\tau_h^{m*}$ along the centres of the CVs just above $y=0.75$. Again the truncation error reduces at a 2nd-order rate, and the quantity $|(\tau_h^m-\tau_h^{m*})/\tau_h^m|$ reduces at a rate which is a little faster than 1st-order. In this case the estimate using (6.5) behaves as good as that using (6.6), and the two estimates are nearly indistinguishable.

Figure 11 displays similar data but for the non-uniform grids. This time $(\tau_h^x-\tau_h^{x*})/\tau_h^x$ does not converge to zero if (6.5) is used, while if (6.6) is used then it decreases again at a rate which is just above 1st-order. Oscillations which appear in the graph of $(\tau_h^x-\tau_h^{x*})/\tau_h^x$ have a period equal to twice the grid spacing of the fine grid $h$, and so they must be due to the prolongation operator (6.7). The behaviour of (6.5) is better for $\tau_h^{m*}$ than for $\tau_h^{x*}$.

Assuming that $(\tau_h-\tau_h^*)/\tau_h \in O(h)$, as numerical experiments confirm when (6.6) is used, then $\tau_h = \tau_h^* + \tau_h \cdot O(h) = \tau_h^* + O(h^{p+1})$, where $\tau_h \in O(h^p)$. It follows that if instead of completely dropping $\tau_h$ in (2.2) one substitutes it by $\tau_h^*$, then the order of the approximation would increase from $p$ to $p+1$. This is confirmed by numerical experiments: When $-\tau_h^{x*}$, $-\tau_h^{y*}$, $-\tau_h^{m*}$ were substituted instead of zero in the right hand sides of the discrete Navier-Stokes and continuity equations respectively, the discretisation errors reduced at a rate which is a little faster than 3rd-order, dropping by nearly an order of magnitude per grid. This is demonstrated in figures 7 and 8 (right). The benefits appear smaller for non-uniform grids, but the distance between the distributions shows 3rd-order accuracy also in this case. This is verified also from figure 12, which shows the magnitude of the approximate integrals of the discretisation errors over the computational domain. The penalty is that the equations must be solved twice on each grid. Similar investigations were conducted in [8].

### *6.2 Flow between concentric cylinders*

Next we consider flow between two concentric cylinders, using a series of grids which exhibit non-orthogonality, expansion and skewness. The inner cylinder has a radius $R_1 = 0.5$ m and is still, and the outer cylinder of radius $R_2 = 1$ m rotates clockwise with a tangential velocity of $V_2 = 1$ m/s (angular velocity $\omega_2 = 1$ rad/s). If we use polar coordinates $r, \phi$ with the angle $\phi$ measured clockwise from the vertical axis $Oy$, then the solution to this problem is a velocity field with a zero component in the $r$-direction and magnitude of $V(r)=Ar+B/r$ where $A=4/3$, $B=-1/3$. The pressure, up to a constant, is given by $p = \rho[A^2r^2/2 - B^2/(2r^2) + 2AB\ln r]$ – see [21]. The fluid has $\rho = 1$ kg/m$^3$, $\mu = 0.01$ Pa·s (viscosity does not affect the solution, but it does affect the magnitude of the truncation error).

Structured grids are used: One family of grid lines consists of straight lines connecting the two cylinders, making a 45° angle with the radial direction at the inner cylinder (henceforth "straight" grid lines). The other family consists of concentric circles, similar to the cylinders themselves (henceforth "circular" grid lines). The finest grid has 512×128 CVs and the circular grid lines lie at radial positions $r_j = 0.75 - 0.25\cos\theta_j$ where $\theta_j = \pi \cdot (j-1)/128$, $j=1,2,\ldots,129$. The other grids come from removing every second grid line from the immediately finer grid. The 64×16 grid is shown in figure 5.

Figure 13 shows $\tau_h^x$ and $\tau_h^m$ on the finest grid. The distribution of $\tau_h^y$ is the same as $\tau_h^x$, rotated by 90°, and $\tau_h^m$ is a function of $r$ only. Again the results are of 2nd-order accuracy as figure 14 shows. The behaviour of the truncation error estimates is similar to the example of section 6.1.2: The estimate based on (6.5) converges to the exact error (in the sense that the ratio $(\tau_h-\tau_h^*)/\tau_h \to 0$) at a rate which is slower than 1st-order, or does not converge at all, while the estimate based on (6.6) converges faster than 1st-order, nearly 2nd-order in some cases. This is demonstrated in figure 15.

This time, if one tries to increase the accuracy by adding $-\tau_h^{x*}$, $-\tau_h^{y*}$, $-\tau_h^{m*}$ to the right hand sides of the Navier-Stokes equations, no solution may be obtained and the iterative procedure does not converge. In fact this is due to the modification of the continuity equation, which may result in the system having no solution. The modified discrete continuity equation for a CV $P$ is:

$$\sum_{f \in f_P} F_{h,f} = -\tau_{h,P}^{m*} \cdot \Delta\Omega_P \qquad (6.8)$$

Each mass flux $F_{h,f}$ through a face of $P$ also appears with opposite sign in the continuity equation of the neighbour CV which also owns face $f$, unless $f$ is a boundary face. Therefore the sum of the left hand sides of (6.8) over all CVs of the grid equals the sum of mass fluxes through the boundary faces, because the mass fluxes through interior faces cancel out. This is zero for the present problem because all boundaries are solid walls. Therefore the sum of the right hand sides of (6.8) should also be zero, but this is not guaranteed by the present method as described so far.

The sum $\sum_P \tau^m_{h,P} \cdot \Delta\Omega_P$ equals zero because the truncation error of the approximation of the mass flux through a face $f$ contributes with opposite sign to the continuity truncation errors of the CVs on either side of $f$. Also, the sum $\sum_P \tau^{h,m}_{2h,P} \cdot \Delta\Omega_P$ over all CVs $P$ of the coarse grid $2h$ is zero because the mass fluxes $F_{2h}$ which are calculated from the restricted velocity and pressure fields contribute with opposite sign to the relative truncation errors of the CVs on either side of the face. This property is lost in converting the relative truncation error on grid $2h$ into a truncation error estimate on grid $h$ by (3.12) - (6.7). Replacing the prolongation operator (6.7) by the operator $(I^h_{2h}\phi_{2h})_C = \phi_{2h,P}$ does not necessarily correct the problem because in the presence of grid skewness the parent CV of grid $2h$ may not cover the same area as its children of grid $h$. However the problem may be overcome by a simple modification of the prolongation operator: If $P$ is a CV of grid $2h$ and $C \in C_P$ where $C_P$ is the set of the children of $P$ then $(I^h_{2h}\tau^{h,m}_{2h})_C$ is multiplied by a function $a_P^{\tau m}$ defined by:

$$a_P^{\tau m} = \frac{\tau^{h,m}_{2h,P} \cdot \Delta\Omega_P}{\sum_{C \in C_P} \left(I^h_{2h}\tau^{h,m}_{2h}\right)_C \cdot \Delta\Omega_C} \qquad (6.9)$$

which ensures that:

$$\sum_{C \in V_h} \tau^{m*}_{h,C} \cdot \Delta\Omega_C = \sum_{C \in V_h} \frac{a_P^{\tau m} \cdot \left(I^h_{2h}\tau^{h,m}_{2h}\right)_C \cdot \Delta\Omega_C}{2^p - 1}$$

$$= \frac{1}{2^p - 1} \sum_{P \in V_{2h}} \left[ a_P^{\tau m} \sum_{C \in C_P} \left(I^h_{2h}\tau^{h,m}_{2h}\right)_C \cdot \Delta\Omega_C \right] = \frac{1}{2^p - 1} \sum_{P \in V_{2h}} \tau^{h,m}_{2h,P} \cdot \Delta\Omega_P = 0$$

where the first equality comes from (3.12) with $I^h_{2h}$ replaced by $a_P^{\tau m} \cdot I^h_{2h}$, and the third equality comes from (6.9). On smooth structured grids (i.e. which are constructed from distributions of dimensionless variables $\xi$, $\eta$ which have continuous derivatives) skewness and expansion tend to zero with grid refinement, so $a_P^{\tau m} \to 1$, and in fact it does so quite rapidly as figure 14 (right) shows. The prolongation operator need not be modified for $\tau^{x*}_h$, $\tau^{y*}_h$. With this modification it was possible to solve the system and to obtain a solution of higher yet still 2nd-order accuracy, as demonstrated in figure 14 (left). We were not able to propose a definite explanation of why 3rd-order accuracy is not achieved, but it is likely that this is due to the boundary conditions. Indeed in [11] it is shown that simple boundary conditions such as those used for the present problems, which are similar to those proposed in [12], result in $\tau^x_h$, $\tau^y_h \in O(1)$ at the boundary CVs. This can be seen also in figure 15, for $\tau^y_h$ at $r = 1$.

## 7. Conclusions

With the aid of a new momentum interpolation, a truncation error estimate has been implemented and tested on smooth structured grids which exhibit non-orthogonality, skewness and stretching. Under these conditions the estimate converges to the exact truncation error in the sense that $(\tau_h - \tau^*_h)/\tau_h \in O(h)$, provided that it uses a restriction operator of sufficient accuracy. In this case the estimate may be used to increase the approximation order of the discrete system, if the boundary conditions are chosen appropriately.

FIGURES

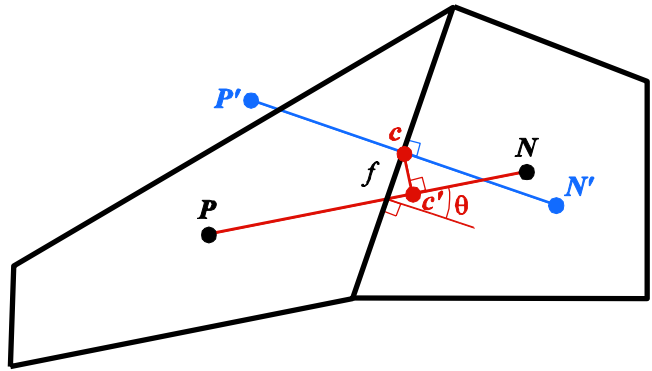

**Figure 1:** Geometry around a face *f* separating control volumes *P* and *N*.

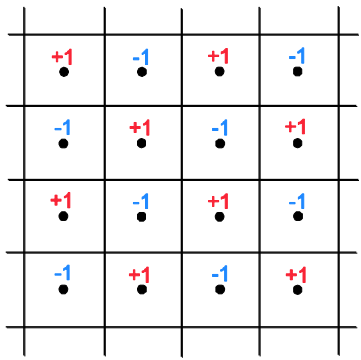

**Figure 2:** Checkerboard distribution of a variable on a Cartesian grid.

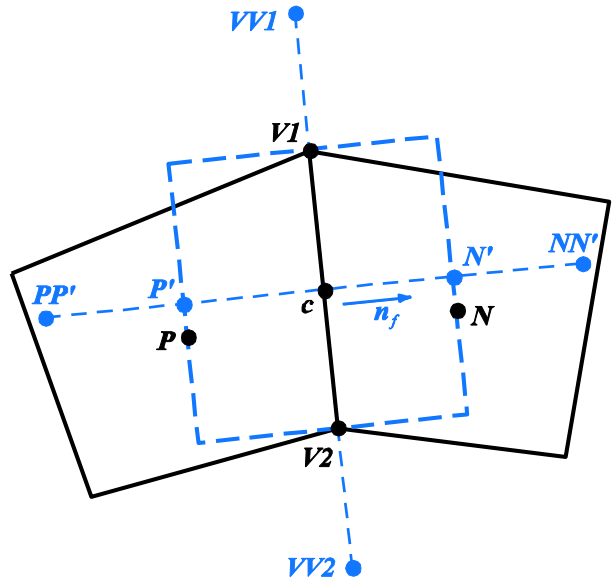

**Figure 3:** Imaginary control volume around a face and related notation.

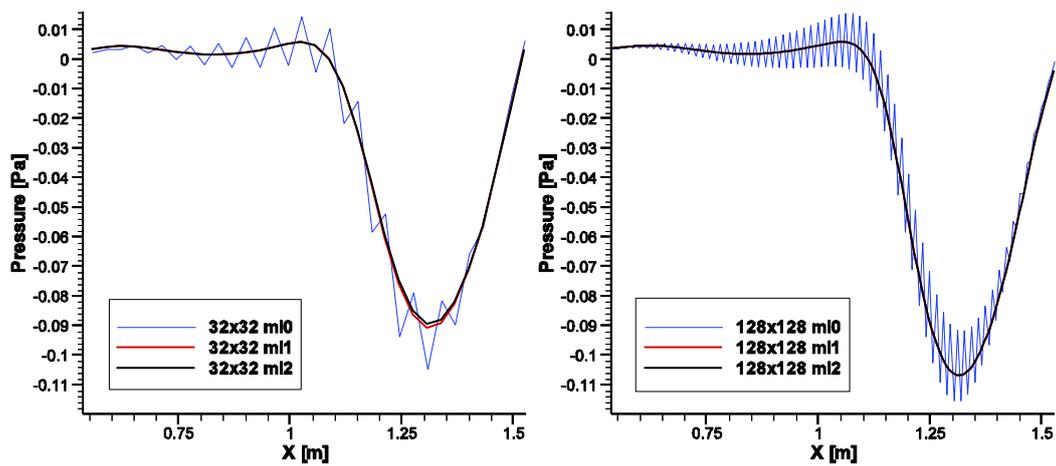

**Figure 4:** 45° skew cavity, Re=1000: Pressure distribution along the CV centres just above the line $y$=3H/4 (H = height of cavity), on grids 32×32 (left) and 128×128 (right).

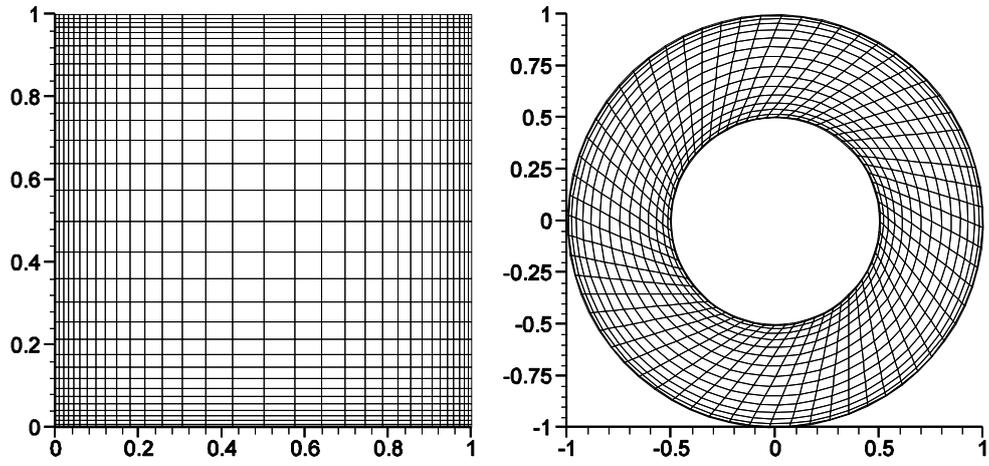

**Figure 5:** *Left*: the 32×32 CV non-uniform grid for the lid-driven cavity problem. *Right*: the 64×16 CV grid for the concentric cylinders problem.

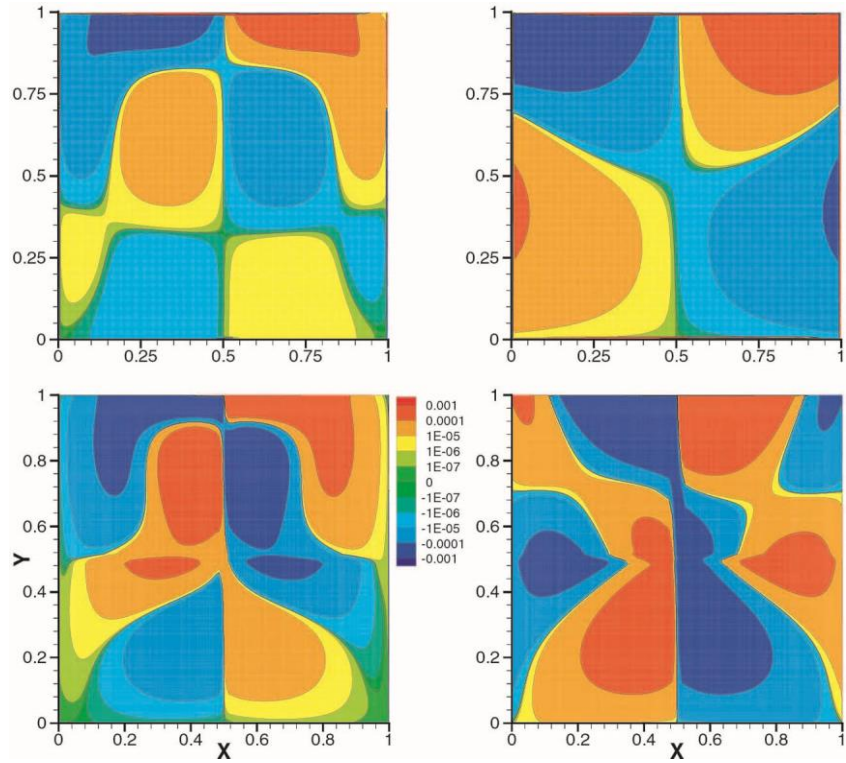

**Figure 6:** The distributions of $-\tau_h^x$ (left) and $-\tau_h^m$ (right) on the 256×256 uniform grid (top) and non-uniform grid (below).

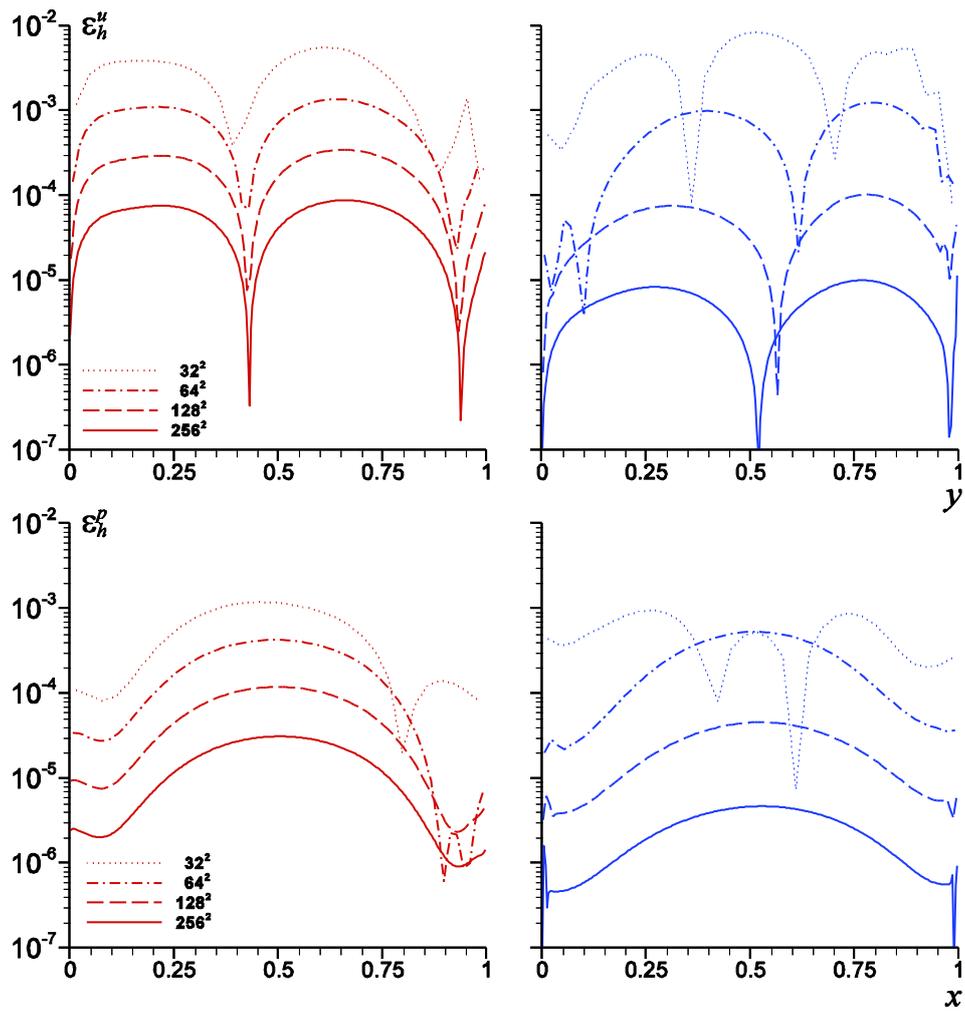

**Figure 7:** *Top:* Distributions of $|\varepsilon_h^u|$ along CV centres just to the right of *x*=0.5, of the solution of the discrete Navier-Stokes system whose right hand side equals zero (left – in red), or equals minus the truncation error estimate (right – in blue), on various uniform grids. *Below:* Similarly for $\varepsilon_h^p$ along CV centres just above *y* = 0.5.

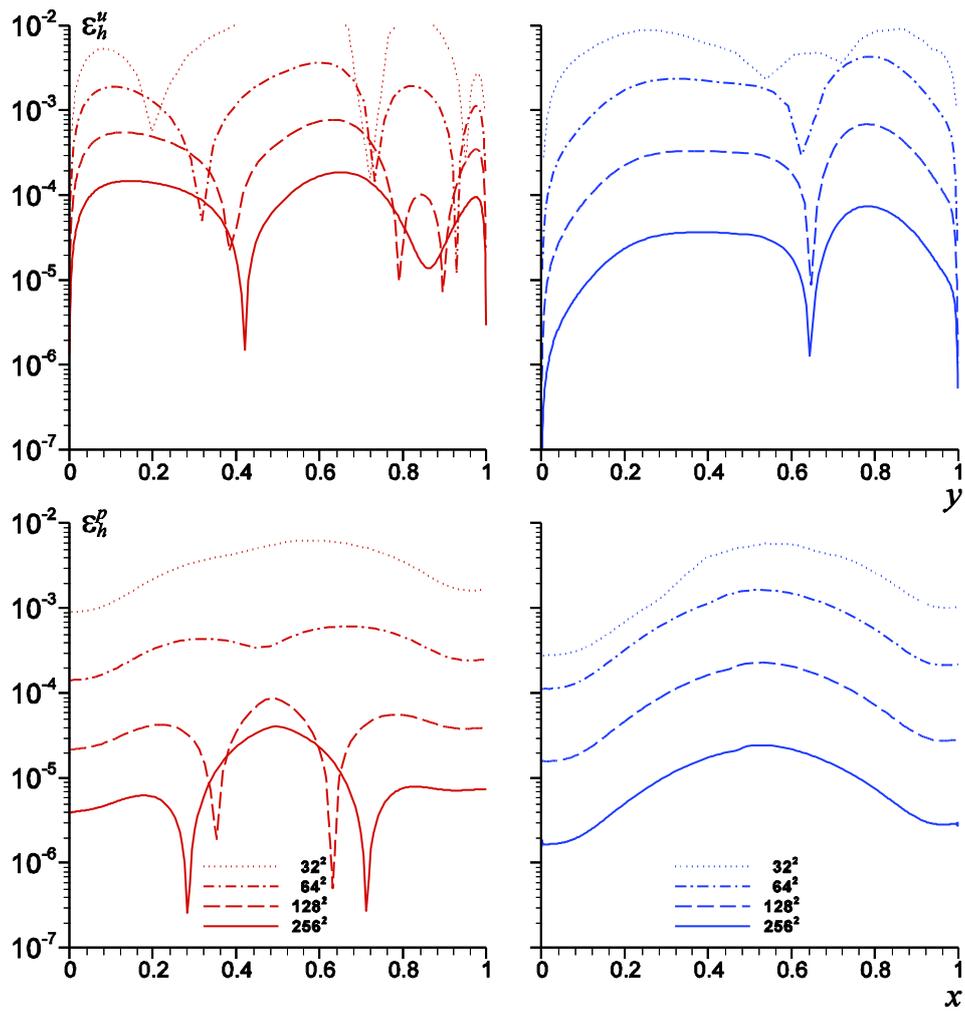

**Figure 8:** Like figure 7, but for solutions on non-uniform grids.

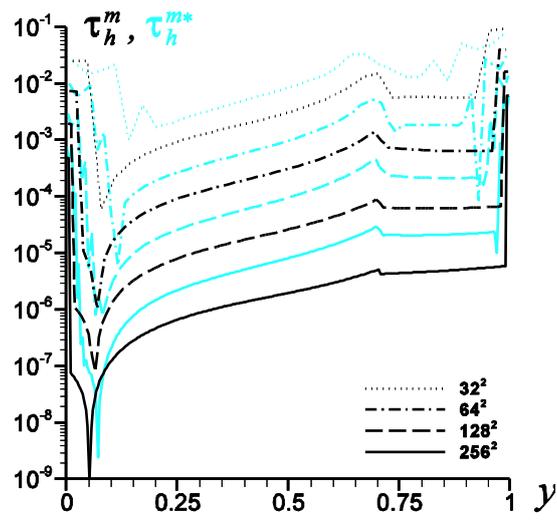

**Figure 9:** Absolute value of $\tau_h^m$ (black) and $\tau_h^{m*}$ (colour) along the CV centres just to the right of $x = 0.5$, for various uniform grids.

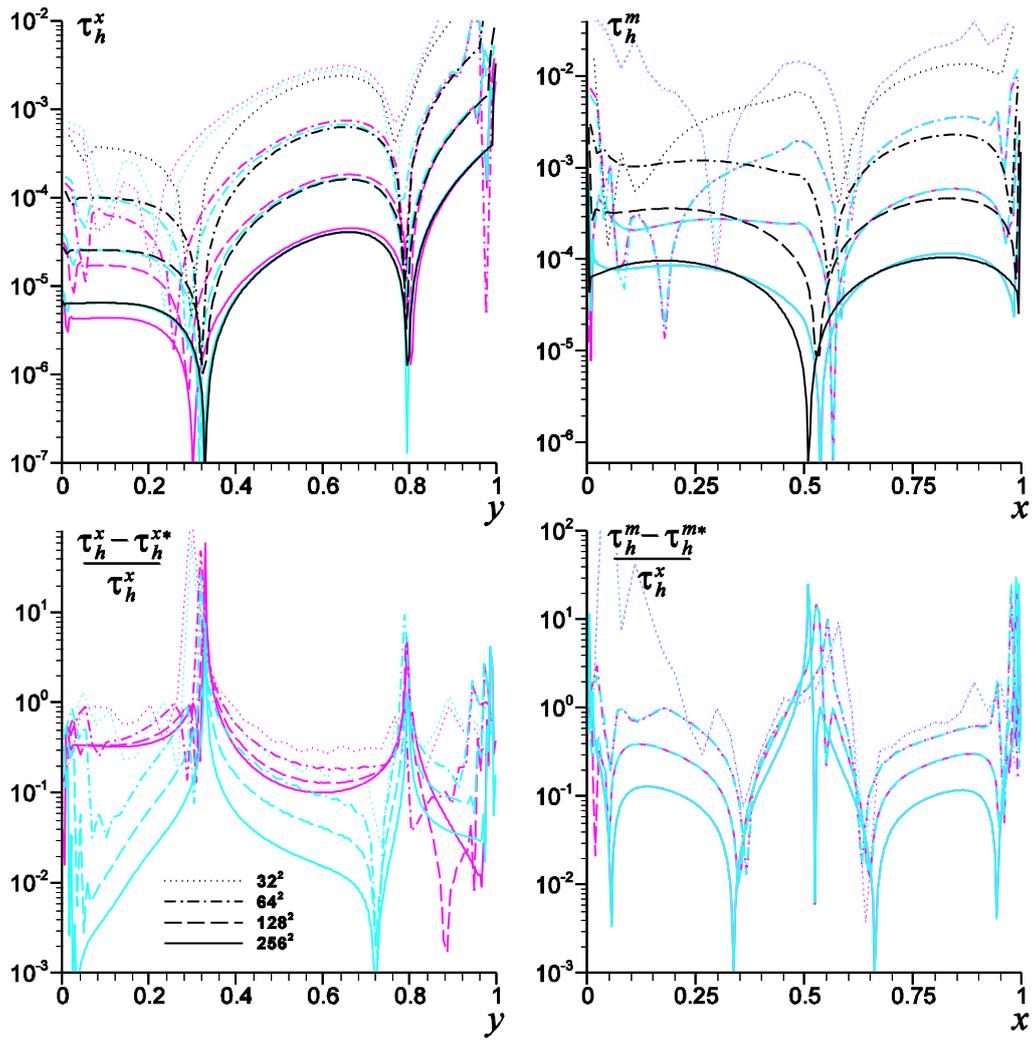

**Figure 10:** *Left side*: The top diagram shows the distributions of $|\tau_h^x|$ (black) and $|\tau_h^{x*}|$ (purple: using (6.5), cyan: using (6.6)) at the CV centres just to the right of $x=0.75$, on various uniform grids, and the bottom diagram shows the corresponding quantities $|(\tau_h^x - \tau_h^{x*})/\tau_h^x|$. *Right side*: Similar to the left side, but for $|\tau_h^m|$, $|\tau_h^{m*}|$ at CV centres just above $y = 0.75$.

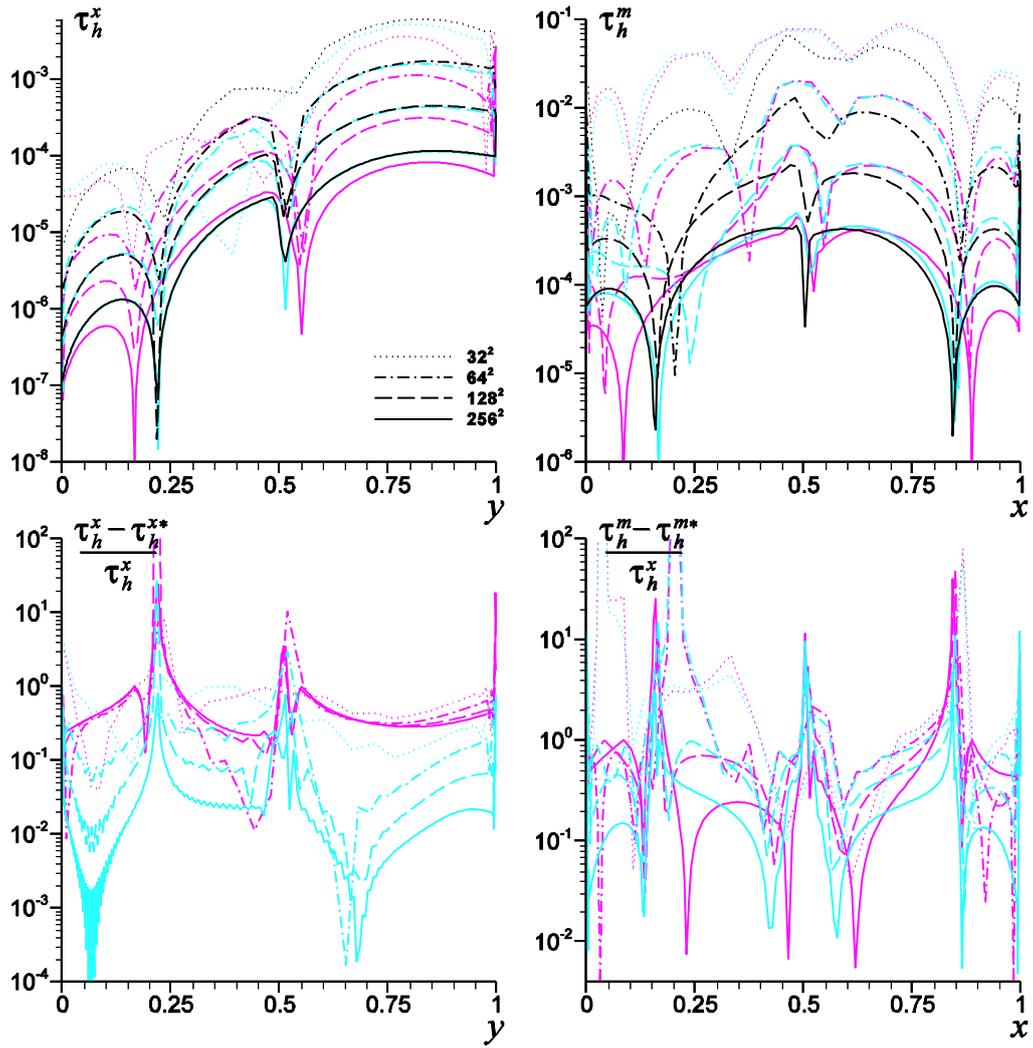

**Figure 11:** *Left side*: Top: $|\tau_h^x|$ (black) and $|\tau_h^{x*}|$ (purple: (6.5), cyan: (6.6)) at CV centres just to the right of the $i=4$ grid line of the 4×4 CV non-uniform grid ($x\approx 0.88$), on various non-uniform grids; bottom: the corresponding quantities $|(\tau_h^x - \tau_h^{x*})/\tau_h^x|$. *Right side*: Similar to the left side, but for $|\tau_h^m|$, $|\tau_h^{m*}|$ at CV centres just above the $j=4$ grid line of the 4×4 CV non-uniform grid ($y\approx 0.88$).

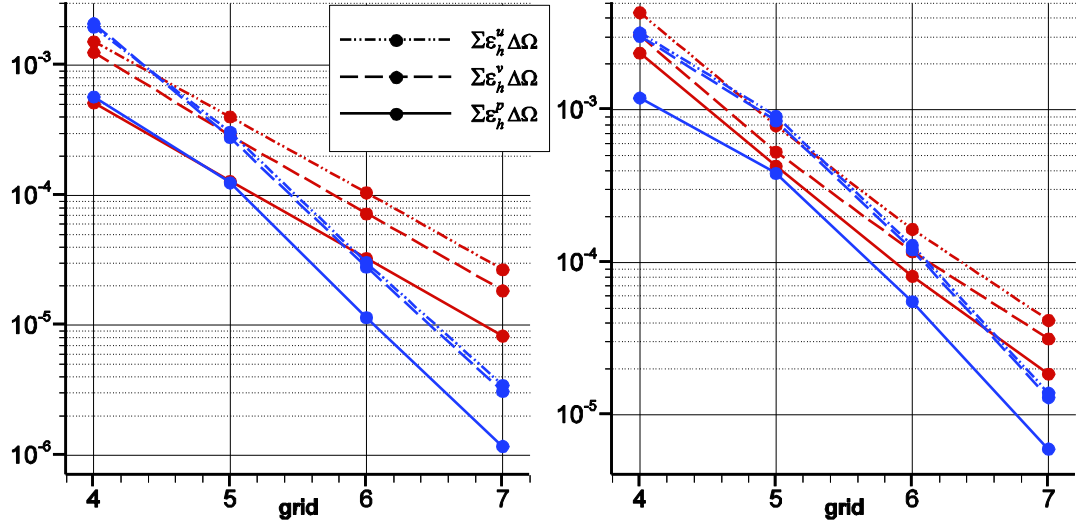

**Figure 12:** $\Sigma_P|\varepsilon^u_{h,P}|\cdot\Delta\Omega_P$, $\Sigma_P|\varepsilon^v_{h,P}|\cdot\Delta\Omega_P$, $\Sigma_P|\varepsilon^p_{h,P}|\cdot\Delta\Omega_P$ of the solution of the original system (red lines) and of the modified system with $-\tau_h^{x*}$, $-\tau_h^{y*}$, $-\tau_h^{m*}$ added to the right hand side (blue lines). The left diagram refers to the series of uniform grids (grid 4 = 32×32 CVs, grid 7 = 256×256 CVs), and the right diagram to the series of non-uniform grids.

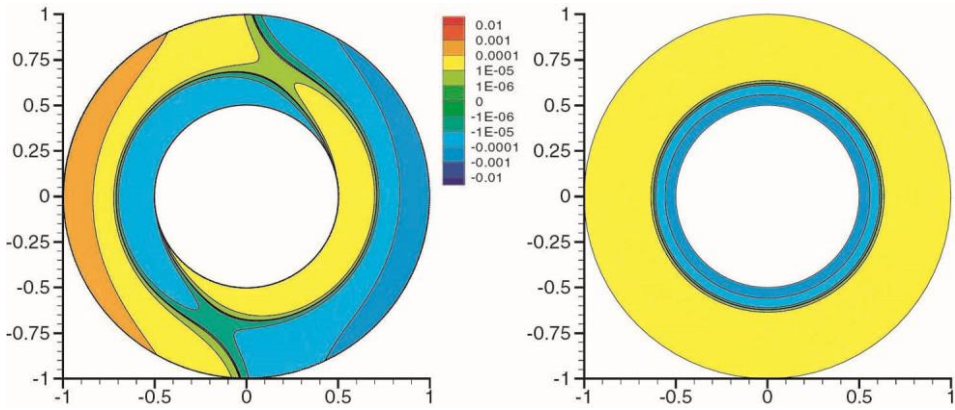

**Figure 13:** $-\tau_h^x$ (left) and $-\tau_h^m$ (right) on the 512×128 grid.

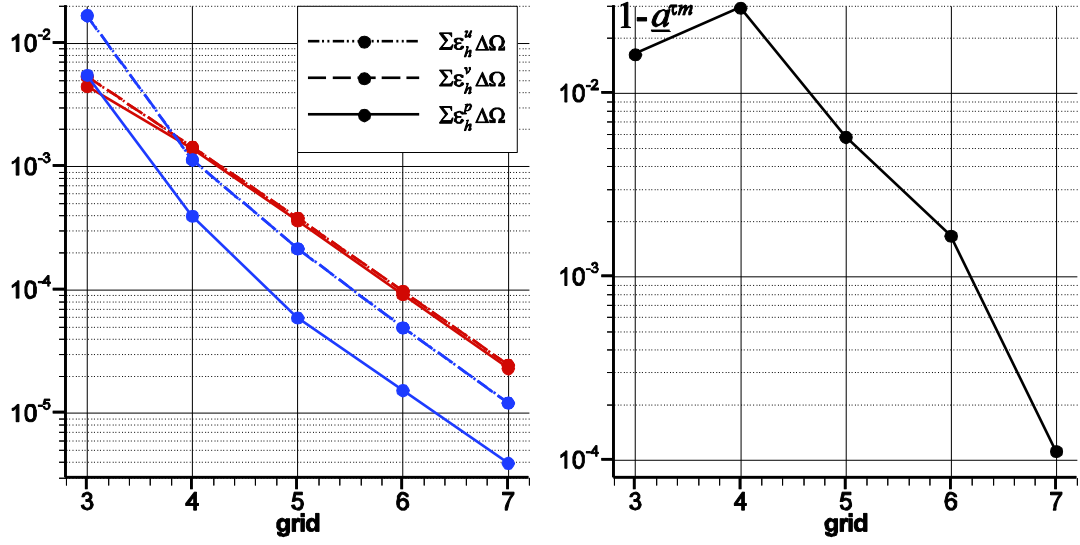

**Figure 14:** *Left*: $\Sigma_P|\varepsilon^u_{h,P}|\cdot\Delta\Omega_P$, $\Sigma_P|\varepsilon^v_{h,P}|\cdot\Delta\Omega_P$, $\Sigma_P|\varepsilon^p_{h,P}|\cdot\Delta\Omega_P$ of the solution of the original system (red lines) and of the modified system with $-\tau_h^{x*}$, $-\tau_h^{y*}$, $-\tau_h^{m*}$ added to the right hand side (blue lines), for the concentric cylinders problem (grid 3 = 32×8 CVs, grid 7 = 512×128 CVs). *Right*: $|1-\underline{a}^{\tau m}|$ where $\underline{a}^{\tau m}$ is the mean value of $|a^{\tau m}|$ (6.9) in the domain.

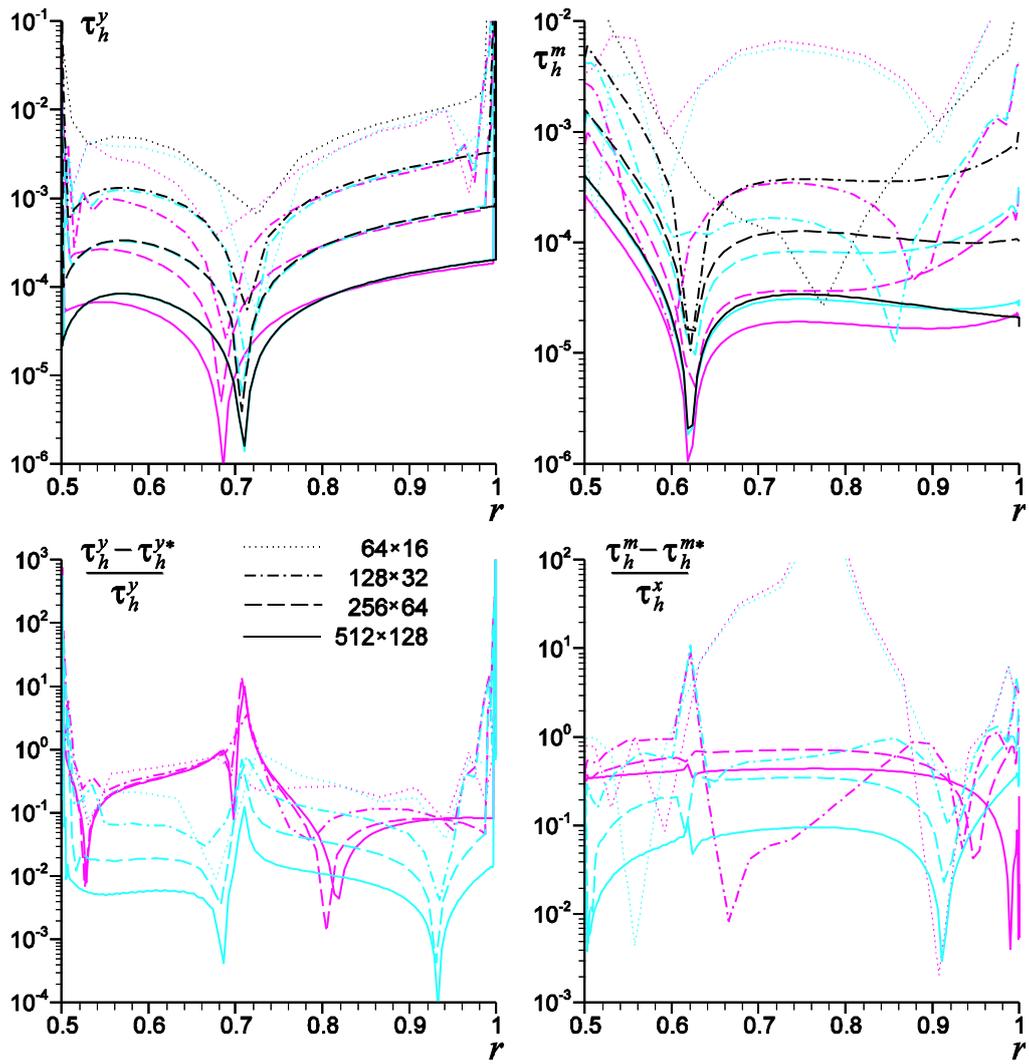

**Figure 15:** $\tau_h^x$, $\tau_h^{x*}$ (left) and $\tau_h^m$, $\tau_h^{m*}$ (right) at CVs immediately to the right of the straight grid line which starts at $(x,y) = (0,0.5)$. The estimate using (6.6) is shown in cyan, and the one using (6.5) in purple.